\documentstyle[12pt]{article}
\newcommand\be{\begin{equation}}
\newcommand\ee{\end{equation}}
\newcommand\bea{\begin{eqnarray}}
\newcommand\eea{\end{eqnarray}}

\newcommand\lt{\left}
\newcommand\rt{\right}

\newcommand\eps{\epsilon}

\newcommand\GeV{{\rm GeV}}

\newcommand\Msusy{M_{\rm s}}

\newcommand\mpl{M_{\rm Pl}}

\newcommand\lsim{\mathrel{\rlap{\lower4pt\hbox{\hskip1pt$\sim$}}
        \raise1pt\hbox{$<$}}}
\newcommand\gsim{\mathrel{\rlap{\lower4pt\hbox{\hskip1pt$\sim$}}
        \raise1pt\hbox{$>$}}}

\newcommand\eq[1]{Eq.~(\ref{#1})}
\newcommand\eqs[2]{Eqs.~(\ref{#1}) and~(\ref{#2})}

\newcommand\bi{\bibitem}
\newcommand\pl[3]{Phys. Lett. #1 (19#3) #2}

\newcommand\pr[3]{Phys. Rep. #1 (19#3) #2}
\newcommand\prl[3]{Phys. Rev. Lett. #1 (19#3) #2}
\newcommand\prd[3]{Phys. Rev. D #1 (19#3) #2}

\begin{document}

\title{Flattening the Inflaton's Potential with \\ Quantum Corrections}
\author{Ewan D. Stewart \\ Research Center for the Early Universe \\
University of Tokyo \\ Tokyo 113, Japan}
\maketitle
\begin{abstract}
I show that a classical scalar potential with $ V''/V \sim 1 $ can be
sufficiently flattened by quantum corrections to give rise to
slow-roll inflation.
This provides perhaps the simplest way to generate an inflationary
potential without fine tuning.
The most natural implementation of this idea produces an unviably
small spectral index, but, for example, $ n \sim 0.8 $ can be obtained
in other implementations.
\end{abstract}
\vspace*{-78ex}
\hspace*{\fill}{RESCEU-19/96}
\thispagestyle{empty}
\setcounter{page}{0}
\newpage
\setcounter{page}{1}

\section{Introduction}

Slow-roll inflation \cite{Linde,KT} requires an unusually flat scalar
potential.
This is quantified by the conditions \footnote{
I set $ \mpl = 1 / \sqrt{8\pi G} = 1 $.}
\be
\lt( \frac{V'}{V} \rt)^2 \ll 1
\ee
and
\be
\label{src2}
\lt| \frac{V''}{V} \rt| \ll 1 \,.
\ee
The first condition is generally not difficult to achieve, but the
second is more problematical \cite{iss}.
In particular, it is straightforward to show that in supergravity
\cite{susy} one generically gets
\be
\lt| \frac{V''}{V} \rt| \gsim 1 \,.
\ee
Most models of inflation resolve this problem by fine tuning.
The only existing proposals for achieving \eq{src2} without fine
tuning were made in \cite{iss,Murayama}.
\footnote{Natural inflation \cite{nat} naturally achieves a small
$V''$ by assuming an approximate global $U(1)$ symmetry, but does
{\em not\/} naturally satisfy \eq{src2} because $V$ vanishes in the
limit where the symmetry is exact. A hybrid natural inflation model
might avoid this problem though \cite{hybrid}.}
The purpose of this paper is to investigate another simpler method.

\section{The idea}
\label{idea}

Any inflationary model must have a non-zero scalar potential energy
density
\be
V = V_0 \,.
\ee
In supergravity this will induce soft supersymmetry-breaking masses
squared of order $V_0$ for all scalar fields\footnote{
We assume that there are no other relevant larger masses, and so
in particular that $ V_0 \gsim \Msusy^4 $ where $ \Msusy $ is the
scale of supersymmetry breaking not due to $V_0$.
If $\Msusy$ takes the same value as in our vacuum then we require
$ V_0^{1/4} \gsim 10^{10} $ to $ 10^{11}\GeV \sim 10^{-8} $ with
$ V_0^{1/4} \sim 10^{-8} $ being the most obvious choice.}
\footnote{For simplicity we take the scalar fields to be real.}
\be
\label{class}
V(\phi) = V_0 \lt[ 1 - \frac{1}{2} A \phi^2 + \ldots \rt]
\ee
with $ |A| \sim 1 $.
The dots represent higher order terms which, in the case of $A>0$,
might naturally stabilise the potential at $\phi \sim 1$.
\eq{class} is our classical potential.
It does not give rise to slow-roll inflation.
In particular, $ |V''/V| \simeq |A| \sim 1 $.

If $\phi$ has either gauge or Yukawa couplings, to vector or chiral
superfields with soft supersymmetry-breaking masses squared of order
$V_0$, quantum corrections will renormalise $\phi$'s mass leading to
a one-loop renormalisation group \cite{susy} effective potential of
the form
\be
\label{veff}
V(\phi) = V_0 \lt[ 1 - \frac{1}{2} f(\eps\ln\phi)
\phi^2 \rt]
\ee
where $ \eps \ll 1 $ is the one-loop suppression factor,
$ f(0) = A + {\cal O}(\eps) $
and the expression is valid for $ V_0 \ll \phi^2 \ll 1 $.
The reader may care to consider $ f(x) = A + B x $ as a toy model.
A more realistic form is given in the Appendix.

Now
\be
\label{v}
V / V_0 = 1 - \frac{1}{2} f \phi^2 \,,
\ee
\be
\label{vp}
V' / V_0 = - \lt( f + \frac{\eps}{2} f' \rt) \phi \,,
\ee
\be
\label{vpp}
V'' / V_0 =
- \lt( f + \frac{3\eps}{2} f' + \frac{\eps^2}{2} f'' \rt) \,.
\ee
Define $\phi_*$ by
\be
\label{star}
f_* + \frac{\eps}{2} f'_* = 0
\ee
where $ f_* \equiv f(\eps\ln\phi_*) $ etc.
Then we can rewrite Eqs.~(\ref{v}),(\ref{vp}) and~(\ref{vpp}) as 
\be
V / V_0 = 1 - \frac{1}{2} \lt[
\eps f'_* \lt( \ln \frac{\phi}{\phi_*} - \frac{1}{2} \rt)
+ {\cal O} \lt( \eps^2 \ln^2 \frac{\phi}{\phi_*} \rt)
\rt] \phi^2 \,,
\ee
\be
\label{vps}
V' / V_0 = - \lt[ \eps \lt( f'_* + \frac{\eps}{2} f''_* \rt)
\ln \frac{\phi}{\phi_*}
+ {\cal O} \lt( \eps^2 \ln^2 \frac{\phi}{\phi_*} \rt)
\rt] \phi \,,
\ee
\be
\label{vpps}
V'' / V_0 = - \eps \lt( f'_* + \frac{\eps}{2} f''_* \rt) - \eps
\lt( f'_* + \frac{3\eps}{2} f''_* + \frac{\eps^2}{2} f'''_* \rt)
\ln \frac{\phi}{\phi_*}
+ {\cal O} \lt( \eps^2 \ln^2 \frac{\phi}{\phi_*} \rt) \,.
\ee
Note that although $ | V''/V | \sim 1 $ over most of the potential,
the quantum corrections have flattened the potential in the vicinity
of $ \phi_* $.

I assume that there exists one, and for simplicity only one, such
$\phi_*$ in the range $ V_0^{1/2} \ll \phi_* \ll 1 $.
We can remove the ambiguity in the definition of $f$ and $\eps$ by
setting
\be
\label{ps}
\eps \ln \phi_* = -1 \,.
\ee
To get $ \phi_* \gsim V_0^{1/2} \gsim 10^{-16} \simeq e^{-37} $
then requires $ \eps \gsim 1/37 $.
We see that the inflaton $\phi$ should have unsuppressed couplings to
other fields, as do for example the Minimal Supersymmetric Standard
Model's up Higgs \cite{susy} or thermal inflation's flaton
\cite{ti,sky}.

\section{Models}
\label{models}

In order for the mechanism of Section~\ref{idea} to work, we require
that \eq{star} has a solution, which for simplicity I assume to be
unique, for $ \phi < 1 $.
This requires that $ f(0) \simeq A $ and $f'_*$ have the same sign.
There are then two classes of potentials depending on whether $f'_*$
is greater or less than zero.
If it is less than zero then $\phi_*$ is a minimum of the potential
and so one must use a hybrid inflation type mechanism \cite{hybrid}
to end inflation at some critical value $\phi_{\rm c}$.
However, as the potential is flat enough for slow roll inflation only
in the neighbourhood of $\phi_*$, one must fine tune to get
$\phi_{\rm c}$ close to $\phi_*$.
We do not consider these models further and from now on assume
$A>0$ and $f'_*>0$.

If $ f'_* > 0 $ then $\phi_*$ is a maximum of the potential.
There are then two possibilities depending on whether $\phi$ rolls
towards the false vacuum at $\phi=0$ or a true vacuum at
$ \phi \sim 1 $.
If $\phi$ rolls towards the false vacuum at $\phi=0$ then one must
again use a hybrid inflation type mechanism to end inflation at some
critical value $\phi_{\rm c}$.
However this time one can take $ \phi_{\rm c} \ll \phi_* $ and so one
does not have to fine tune $\phi_{\rm c}$.
The spectral index of such a model would change from $n<1$ to $n>1$,
possibly over observable scales.
However, it seems difficult to achieve the required initial conditions
of $ \phi \simeq \phi_* $.
The unnaturalness of the initial conditions might though be
compensated to some extent by the eternal inflation \cite{Linde} that
occurs at $ \phi = \phi_* $.

In the case of the slow roll inflation occuring as $\phi$ rolls
towards the true vacuum at $ \phi \sim 1 $, we have the natural
initial condition $ \phi = 0 $.
At first we have (eternal) old inflation with $ \phi = 0 $,
then quantum fluctuations kick $\phi$ over the barrier at
$ \phi = \phi_* $ (or $\phi$ tunnels through the barrier),
and slow roll inflation occurs as $\phi$ rolls off to the new vacuum
at $ \phi \sim 1 $.
In order to obtain a sufficiently high reheat temperature, at least in 
the case of $ V_0 \sim 10^{-32} $, we require that $\phi$ has
new couplings to new superfields which become light at the new vacuum.
This is a natural expectation in string (or M or F) theory.
If $\phi$ corresponded to an Affleck-Dine field in the {\em new\/}
vacuum, it might even be able to generate a large enough baryon number
to survive the diluting effects of thermal inflation as its initial
amplitude would be of the order of the Planck scale.
However, the baryogenesis mechanism of \cite{sky} is more compelling.
This model is theoretically healthy\footnote{
Indeed, it is perhaps the first truly natural and essentially
complete model of slow roll inflation.
Refs.~\cite{iss} and~\cite{Murayama}, although extremely promising and
potentially natural, can not yet be regarded as essentially complete as
they require detailed knowledge of the as yet imprecisely understood
high energy theory.
The present model on the other hand uses $A$ of \eq{class} to
parameterize the unknown high energy physics and depends only on the
qualitative features of the low energy vacuum, albeit a different low
energy vacuum ($\phi=0$) than our low energy vacuum
($ \phi \sim 1 $).}
but, as we shall see in the next section, tends to give an
uncomfortably small spectral index especially for the favoured
$ V_0 \sim 10^{-32} $.

\section{The spectral index}

From \eq{vp} the equation of motion for $\phi$ is
\be
\ddot{\phi} + 3H \dot{\phi}
- V_0 \lt( f + \frac{\eps}{2} f' \rt) \phi = 0 \,.
\ee
Define $ 3H_0^2 = V_0 $, $ x = \eps\ln\phi $ and
$ \tau = \eps H_0 t $, and take a dot to now represent differentiation
with respect to $\tau$ rather than $t$ which it did before. Then,
assuming $ H \simeq H_0 $ which will be a good approximation for
$ \phi \ll 1 $, we get
\be
\eps \ddot{x} + \dot{x}^2 + 3\dot{x} - 3f - \frac{3\eps}{2} f' = 0 \,.
\ee
Neglecting the first term and taking the branch that connects to the
slow roll trajectory gives
\be
\label{dif}
\dot{x} = - \frac{3}{2} \lt( 1 -
\sqrt{ 1 + \frac{4}{3} f + \frac{2\eps}{3} f' } \rt) \,.
\ee
Differentiating gives
\be
\frac{\ddot{x}}{\dot{x}} = \frac{ f' + \frac{\eps}{2} f'' }{
\sqrt{ 1 + \frac{4}{3} f + \frac{2\eps}{3} f' } }
\ee
and so the error in \eq{dif} is of order $\eps$ unless the square
root becomes small.
Integrating \eq{dif} gives
\be
\label{int}
\tau = - \frac{2}{3} \int \frac{dx}{1 -
\sqrt{ 1 + \frac{4}{3} f + \frac{2\eps}{3} f' } } \,.
\ee
The slow roll approximation corresponds to expanding in $ x - x_* $.
Working to lowest order in $\eps$ and second order in the slow roll
approximation and using \eq{star} gives
\be
1 - \sqrt{ 1 + \frac{4}{3} f + \frac{2\eps}{3} f' }
= - \frac{2}{3} f'_* \lt( x - x_* \rt)
\lt[ 1 + \frac{f'_*}{3} \lt( \frac{3f''_*}{2{f'_*}^2} - 1 \rt)
\lt( x - x_* \rt) \rt]
\ee
and so
\be
\label{tau}
\tau = \frac{1}{f'_*} \ln \lt( x - x_* \rt)
- \frac{1}{3} \lt( \frac{3f''_*}{2{f'_*}^2} - 1 \rt) \lt( x - x_* \rt)
\ee
where the second term is of second order in the slow roll approximation.
The number of $e$-folds to the end of inflation is given by
$ N = ( \tau_2 - \tau ) / \eps $ where subscript 2 denotes the end of
inflation.
Rewriting \eq{tau} and working to lowest order in the slow roll
approximation at $\phi$, but not at $\phi_2$, gives
\be
\label{N}
N = \frac{1}{\eps f'_*} \ln \lt[
\frac{ \ln \lt( \phi_2 / \phi_* \rt) }{ \ln \lt( \phi / \phi_* \rt) }
\rt] - C
\ee
where $C$ is a constant given by
\be
C =  \frac{1}{3} \lt( \frac{3f''_*}{2{f'_*}^2} - 1 \rt)
\ln \frac{ \phi_2 }{ \phi_* }
+ {\cal O} \lt( \eps \ln^2 \frac{ \phi_2 }{ \phi_* } \rt)
\,.
\ee
$C$ is calculated exactly in the Appendix for a semi-realistic example 
and found to be given by $ C = 2/3\eps $ for $ \phi_2 \sim 1 $.
Inverting \eq{N} and using \eq{ps} gives
\be
\label{phi}
\ln \frac{ \phi }{ \phi_* } = \ln \frac{ \phi_2 }{ \phi_* }
e^{ - \eps f'_* (N+C) }
= \frac{1}{\eps} \lt( 1 + \eps \ln \phi_2 \rt)
e^{ - \eps f'_* (N+C) } \,.
\ee

The COBE normalisation and \eq{vps} give
\be
\frac{ V_1^{3/2} }{ V'_1 } =
\frac{ V_0^{1/2} }{ \eps f'_* \phi_1 \ln ( \phi_1 / \phi_* ) }
= 6 \times 10^{-4}
\ee
where subscript 1 denotes the time when COBE scales left the horizon. 
For $ \eps f'_* \ln ( \phi_1 / \phi_* ) \sim 10^{-1} $ to $ 10^{-2} $
and $ V_0 \sim 10^{-32} $ this gives
$ \phi_1 \sim 10^{-11} \sim e^{-25} $.
Slow roll at $\phi_1$ requires
$ \eps \ln ( \phi_1 / \phi_* ) \ll 1 $ and so we get
$ \eps \simeq 0.04 $.
More generally, using \eqs{ps}{phi}, it gives
\be
V_0^{1/2} = f'_* \lt( 1 + \eps \ln \phi_2 \rt) 
\exp \lt[ - \frac{1}{\eps} - 7.4 + \frac{1}{\eps}
\lt( 1 + \eps \ln \phi_2 \rt) e^{ - \eps f'_* (N_1+C) }
- \eps f'_* (N_1+C) \rt] \,.
\ee
The spectral index is, using \eqs{vpps}{phi},
\bea
n & = & 1 + 2 \frac{V''}{V}
= 1 - 2 \eps f'_* \lt( \ln \frac{ \phi }{ \phi_* } + 1 \rt) \\
\label{n}
& = & 1 - 2 f'_* \lt[ \lt( 1 + \eps \ln \phi_2 \rt)
e^{ - \eps f'_* (N+C) } + \eps \rt] \,.
\eea
A natural choice of parameters would be $ \phi_2 \sim 1 $,
$ V_0 \sim 10^{-32} $ and the $f$ of \eq{f} with $A=1$.
Observable scales would then leave the horizon at $ N \sim 30 $,
assuming thermal inflation.
Unfortunately for these parameters, although slow roll does occur for
sufficiently large $N$, it has finished by $ N \sim 30 $ and we get an
unacceptable spectral index.
This situation can be improved in a number of ways:
\begin{enumerate}
\item
Increase $N$ by either abandoning thermal inflation or increasing
$V_0$.
\item
Increase $\eps$ by increasing $V_0$.
\item
Increase $C$ by fiddling with $f$.
\item
Increase $f'_*$ by fiddling with $f$ or $A$.
\item
Decrease $\phi_2$, \mbox{i.e.} end inflation earlier, by for example
using a hybrid inflation type mechanism \cite{hybrid}.
\item
Take $ \phi_2 < \phi_* $ so that a partial cancellation occurs in $n$.
This corresponds to the model of Section~\ref{models} with the
problematic initial conditions.
\end{enumerate}
To illustrate how these effects can help we take $ \phi_2 \sim 1 $,
$ \eps = 0.1 $, $ N = 40 $, $ C = 2/3\eps $ and $ f'_* = 0.5 $
corresponding to $ V_0^{1/4} \sim 10^{14}\,\GeV $, having thermal
inflation, and the $f$ of \eq{f} with $A=1$.
\eq{n} then gives $ n = 0.8 $ which is the spectral index recently
argued for in \cite{Andrew}.

\section{Conclusions}

I have described a way to obtain slow-roll inflation which is
completely natural from the particle physics point of view
and should be realisable in string (or M or F) theory.
The spectral index has the general form
\be
n = 1 - \alpha e^{ - \beta N } - 2 \beta 
\ee
and is likely to show an observable change over observable scales.
Unfortunately, the most natural values of the parameters tend to give
a spectral index that is too small to be consistent with observations.
However, plausible values of the parameters can give the spectral
index of $ n \sim 0.8 $ recently argued for in \cite{Andrew}.

\subsection*{Acknowledgements}
I would like to thank Lisa Randall for stimulating discussions and the 
Aspen Center for Physics for its hospitality when these discussions
took place.
I also thank David Lyth for many helpful discussions.
I am supported by a JSPS Fellowship at RESCEU, and my work is
supported by Monbusho Grant-in-Aid for JSPS Fellows No.\ 95209.

\subsection*{Appendix}

As an example we take $f$ to be given by
\be
\label{f}
f(x) = \frac{3}{4} \lt[ \lt( \frac{a+1}{a-x} \rt)^2 -1 \rt]
\ee
with
\be
a = \frac{3}{4A} \lt( 1 + \sqrt{ 1 + \frac{4A}{3} } \rt) \,.
\ee
This is a realistic form except for the precise value of the
coefficient 3/4 which was chosen to make the integral in \eq{int}
analytically soluble.
It corresponds to a gauge coupling which increases with the energy
scale and negligible Yukawa couplings.
As \eq{dif} has an error of order $\eps$, we work to lowest order in
$\eps$.
We have $f(0)=A$ and $f(-1)=0$ in accord with the definitions of
Section~\ref{idea}.
Also $f'(-1)=(3/2)(a+1)^{-1}$, $f''(-1)=(9/2)(a+1)^{-2}$ and for $A=1$
we have $ f'(-1) \simeq 0.5 $.
Substituting \eq{f} into \eq{int} gives
\be
\tau = \frac{1}{f'(-1)} \ln \lt( 1 + x \rt) - \frac{2}{3} x
\ee
and so for the above choice of $f$ the $C$ of \eq{N} is given by
\be
\label{CA}
C = \frac{2}{3\eps} \lt( 1 + \eps \ln \phi_2 \rt) \,.
\ee
Note that the second order slow roll formula gives the exact result
in this case.

\end{document}